\documentclass[12pt]{myiopart}
\usepackage[fleqn]{amsmath}
\usepackage{amsfonts}
\usepackage{amssymb}
\usepackage{color}

\definecolor{MyDarkBlue}{rgb}{0.15,0.15,0.45}
\definecolor{MyLightBlue}{rgb}{0.15,0.45,0.45}
\usepackage[linktocpage=true]{hyperref}
\hypersetup{
colorlinks=true,
citecolor=MyDarkBlue,
linkcolor=MyDarkBlue,
urlcolor=MyLightBlue,
}

\begin{document}

\title{On one-loop partition functions of three-dimensional critical gravities}

\author{Thomas Zojer}

\address{Centre for Theoretical Physics, University of Groningen,\\
         Nijenborgh 4, 9747 AG Groningen, The Netherlands}
\ead{t.zojer@rug.nl}

\begin{abstract}
 We calculate the one-loop partition function of three-dimensional parity even tricritical gravity. Agreement
 with logarithmic conformal field theory single-particle partition functions on the field theory side is found and we
 furthermore discover a partially massless limit of linearized six-derivative parity even gravity. Then we define a ``truncation''
 of the critical theory, at the level of the partition function, by calculating black hole determinants via summation
 over quasi-normal mode spectra and discriminating against those modes which are not present in the physical spectrum.
 This ``truncation'' is applied to critical new massive gravity and three-dimensional parity even tricritical
 gravity.
\end{abstract}




\section{Introduction}

Three-dimensional gravity has long been an interesting testing ground for theories of quantum gravity. But pure Einstein gravity,
plus a cosmological constant, does not lead to propagating degrees of freedom in the bulk. Therefore modifications
are sought. One natural way of extending the theory is to add massive gravitons. This can be achieved by adding
higher-derivative terms to
the action. Two concrete proposals in the literature are the parity violating topologically massive gravity (TMG)
\cite{Deser:1981wh} with one massive (helicity) degree of freedom, and parity even new massive gravity (NMG)
\cite{Bergshoeff:2009hq} with two massive (helicity) degrees of freedom propagating in a unitary fashion around an anti-de
Sitter (AdS) background. 

Lately, one center of attention in the study of three-dimensional higher-derivative gravity were so-called logarithmic or
critical tunings of the parameters. The seminal example is chiral gravity \cite{Li:2008dq}, a fine-tuned version of TMG,
which was conjectured to be dual to a logarithmic conformal field theory (LCFT) \cite{Grumiller:2008qz}. Besides the
latter example the phenomenon of critical points was also studied in NMG, general massive gravity (GMG)
\cite{Bergshoeff:2009hq,Bergshoeff:2009aq} and more recently in parity even tricritical gravity (PET) \cite{Bergshoeff:2012ev}. For
a summary on the works contributing to the LCFT conjectures in three-dimensional critical gravities see
e.g.~\cite{Bergshoeff:2012ev,Grumiller:2010tj} and references therein.

One piece of evidence for the aforementioned critical gravity/LCFT conjecture was the calculation of one-loop partition
functions on the gravity side and the results were shown to agree with those expected from a LCFT. This computation was performed
for TMG and NMG in \cite{Gaberdiel:2010xv} and for GMG in \cite{Bertin:2011jk}. In this work we calculate the
one-loop partition function of PET gravity along the lines of \cite{Gaberdiel:2010xv}. We will scan the
whole parameter space of PET
gravity, covering the critical points found in \cite{Bergshoeff:2012ev}, but also uncovering an additional special point
in parameter space: the partially massless limit of linearized six-derivative parity even gravity. Focusing on the
critical points we will give further evidence for the conjectured duality of PET gravity to LCFTs of rank two and three.

The AdS/\emph{L}CFT duality only holds for those theories where \emph{all} solutions to the equations of motion are allowed.
From the very beginning probably one of the most prominent questions concerning critical gravities was how (and if) one
can consistently truncate (some of) the logarithmic solutions by imposing boundary conditions. For example restricting
to Brown--Henneaux boundary conditions
\cite{Brown:1986nw} for excitations on the gravity side would kill all log modes. In the prime example of TMG/chiral
gravity these are the boundary conditions one has to choose to obtain chiral gravity. If one imposes log boundary conditions
\cite{Grumiller:2008es} one obtains log-TMG which is dual to a LCFT.
Recently, using a scalar field toy model, another truncation of critical theories was put forward that allows some, but not
all log solutions \cite{Bergshoeff:2012sc}. In \cite{Bergshoeff:2012ev} a tricritical version of this scalar field toy model
was generalized to interacting spin-two fields. Calculations on the linearized level seemed to lend support to the possibility of
truncating the theory. However, it was shown that the theory is flawed with a linearization instability \cite{Apolo:2012he}
and only makes sense either as dual to a LCFT by allowing all solutions, or, imposing Brown--Henneaux boundary conditions,
as a 'trivial' theory propagating null modes.

In the second part of this work we introduce an idea on how such a truncation via boundary conditions could be understood at the
level of the partition function. To do so we will interpret the partition function as a sum over quasi-normal mode frequencies
following the idea of \cite{Denef:2009kn}. The authors of \cite{Denef:2009kn} emphasize that translating from the heat-kernel
to the quasi-normal mode spectrum would allow one to pick out the contribution of only one mode or only one frequency to the partition
function. We try to turn this argument around by using it to dismiss the contributions of certain modes. Our ``truncation'' will
effectively be a prescription on which quasi-normal modes to keep in the spectrum and which not to keep. We obtain results
concurrent with the comment made in the previous paragraph: either we keep all modes and the theory is dual to a non-unitary LCFT,
or it is not logarithmic but an ``ordinary'' CFT that propagates only null modes. 

This work is organized as follows. In section \ref{sec2} we calculate the partition function of PET gravity and discuss
its similarity to LCFT single-particle partition functions. In section \ref{sec3} we define the truncation by translating
from black hole determinants to quasi-normal mode spectra and discriminating against certain modes. Finally we apply this
truncation to critical NMG and PET gravity.

\section{Partition function of PET gravity}\label{sec2}

In this first part of the work we calculate the gravity one-loop partition function of PET gravity. Such a calculation is
conveniently carried out by splitting the metric into a background metric $g_{\mu\nu}$ and a perturbation $h_{\mu\nu}$.
Our background is an AdS space-time, but, as we will argue later, this is equivalent to taking the BTZ black hole
\cite{Banados:1992wn} as background. 
The partition function then consist at least of a classical part, $Z_c$, corresponding to the background $g_{\mu\nu}$, and a
one-loop contribution coming from the perturbation $h_{\mu\nu}$, $Z^{\rm 1-loop}$. This one-loop contribution is precisely
what we are interested in.

We start by covering the whole parameter range of $\beta$ and $b_2$ [see the action (\ref{linaction})], or equivalently the masses
$M_+$ and $M_-$ of the two propagating massive gravitons. Subsequently, we specialize to the logarithmic points and
subsection \ref{critloc} is devoted to the critical loci: the tricritical point, the logarithmic line (single log) where
one of the two masses $M_+$ and $M_-$ goes to zero, and the massive logarithmic line (massive log) where $M_+$ and $M_-$
degenerate with each other. Section \ref{pmPET} covers a special parameter limit that leads to a
partially massless theory. At the critical loci the calculation confirms the conjecture that PET gravity is dual to
parity even logarithmic CFTs of rank two and three. This is shown in section \ref{CFTinterpret} using the combinatorial
counting argument of \cite{Gaberdiel:2010xv}.

The calculation will follow the lead of \cite{Gaberdiel:2010xv} and all formulas and technicalities that are
not explained in detail here can be found in that reference.

\subsection{Linearized action and ghost determinants}

The semi-classical one-loop contribution to the gravity partition function is given by
\begin{equation}\label{pathint}\hskip 1cm
 Z_{\rm PET}^{\rm 1-loop}=\int\mathcal{D}h_{\mu\nu}\mathcal{D}k_{\mu\nu}\mathcal{D}f_{\mu\nu} \,
                          e^{-\delta^{(2)} S_{\rm PET}} \,.
\end{equation}
The action in the exponent is the linearized Euclidean action of PET gravity, given by \cite{Bergshoeff:2012ev}
\begin{equation}\begin{split}\label{linaction}\hskip 1cm
 \delta^{(2)} S_{\rm PET} = \frac{1}{\kappa^2}\int{\rm d} ^3x\sqrt{g} \, \Big\{
        &   -\frac12 \bar{\sigma} h^{\mu\nu}\mathcal{G}_{\mu\nu}(h) +k^{\mu\nu}\mathcal{G}_{\mu\nu}(h)
           +2b_2f^{\mu\nu}\mathcal{G}_{\mu\nu}(f) \\
        &   +(2b_2/\ell^2+\beta)(f^{\mu\nu}f_{\mu\nu}-f^2)-(f^{\mu\nu}k_{\mu\nu}-fk)  \Big\} \,,
\end{split}\end{equation}
with $\bar{\sigma}=\sigma+3b_2/(2\ell^4)+\beta/(2\ell^2)$. Newton's constant $G$ is in $\kappa^2=16\pi G$ and the parameters
$\beta$ and $b_2$ have mass dimension minus two and minus four respectively, but are otherwise arbitrary parameters of the model.
The linearized
Einstein tensor $\mathcal{G}_{\mu\nu}$ --- without imposing any gauge condition --- takes the form\footnote{The symmetrization
over indices is normalized as follows:
$2\nabla_{(\mu}\nabla^\beta h_{\nu)\beta}=\nabla_{\mu}\nabla^\beta h_{\nu\beta}+\nabla_{\nu}\nabla^\beta h_{\mu\beta}$.}
\begin{equation}\hskip 1cm
  2 \mathcal{G}_{\mu\nu}(h) = -\nabla^2h_{\mu\nu}-\nabla_\mu\nabla_\nu h+2\nabla_{(\mu}\nabla^\beta h_{\nu)\beta}
                            -\frac{2}{\ell^2} h_{\mu\nu}-g_{\mu\nu}(\nabla_\rho\nabla_\sigma h^{\rho\sigma}-\nabla^2h) \,.
\end{equation}

All fluctuations $h_{\mu\nu}$, $k_{\mu\nu}$ and $f_{\mu\nu}$ can be split into a transverse-traceless (TT), a
trace, and a vector (or gauge) part:
\begin{equation}\begin{split}\label{tensorsplit}\hskip 1cm
 &h_{\mu\nu}(h^{TT},h,\xi) = h_{\mu\nu}^{TT}+\frac13 g_{\mu\nu}h +\nabla_{(\mu}\xi_{\nu)} \\
 &k_{\mu\nu}(k^{TT},\bar{K},v) = k_{\mu\nu}^{TT}+\frac13 g_{\mu\nu}\bar{K} +\nabla_{(\mu}v_{\nu)} \\
 &f_{\mu\nu}(f^{TT},\bar{F},u) = f_{\mu\nu}^{TT}+\frac13 g_{\mu\nu}\bar{F} +\nabla_{(\mu}u_{\nu)} \,.
\end{split}\end{equation}
Note that by definition $g^{\mu\nu}h_{\mu\nu}^{TT}=\nabla^\mu h_{\mu\nu}^{TT}=0$ and that the vector parts contribute
to the traces of $g^{\mu\nu}k_{\mu\nu}=\bar{K}+2\nabla_\mu v^\mu$ and $g^{\mu\nu}f_{\mu\nu}=\bar{F}+2\nabla_\mu u^\mu$.

Each such decomposition of the fluctuations produces a 'ghost' factor \cite{Gaberdiel:2010xv} in the measure of the
path-integral (\ref{pathint}).
\begin{equation}\hskip 1cm
\mathcal{D}h_{\mu\nu}=Z_{\rm gh}\,\mathcal{D}h_{\mu\nu}^{TT}\mathcal{D}\xi_\mu\mathcal{D}h
\end{equation}

Another useful split is
\begin{equation}\label{vectorsplit}\hskip 1cm
 u^\mu=u^\mu_T + \nabla^\mu \delta \,,
\end{equation}
where $\nabla_\mu u^\mu_T=0$. This yields the ghost factor $J_1$.
\begin{equation}\hskip 1cm
\mathcal{D}u_\mu=J_1\,\mathcal{D}u_\mu^T\mathcal{D}\delta
\end{equation}
The values of $Z_{\rm gh}$ and $J_1$ are given by \cite{Gaberdiel:2010xv}
\begin{equation}\label{Zghost}\hskip 1cm
 Z_{\rm gh}= \big[\det(-\nabla^2+2)_1^T\det(-\nabla^2+3)_0\big]^{1/2} \quad{\rm and}\quad
        J_1= \big[\det(-\nabla^2)_0\big]^{1/2}\,.
\end{equation}

\subsection{Path-integral for PET gravity}

We now consider the action (\ref{linaction}) under the decompositions (\ref{tensorsplit}) and (\ref{vectorsplit}). All scalar,
'vector' (T) and 'tensor' (TT) modes decouple and we can write the action (\ref{linaction}) as a sum of terms which are
quadratic in the respective fluctuations. For the transverse-traceless part we find
\begin{equation}\begin{split}\label{actionTT}\hskip 1cm
 \delta^{(2)}S_{\rm PET}^{TT}= \int {\rm d} ^3x\sqrt{g} \,\Big\{
   &  -\frac{\bar{\sigma}}{4} h_{TT}^{\mu\nu}\big(-\nabla^2-\frac{2}{\ell^2}\big)h_{\mu\nu}^{TT}
      + \frac12 k_{TT}^{\mu\nu}\big(-\nabla^2-\frac{2}{\ell^2}\big)h_{\mu\nu}^{TT} + \\ 
  & + b_2 f_{TT}^{\mu\nu}\big(-\nabla^2+\frac{\beta}{b_2}\big)f_{\mu\nu}^{TT}  -f^{\mu\nu}_{TT}k_{\mu\nu}^{TT}\Big\} \,.
\end{split}\end{equation}
The transverse vector part is given by
\begin{equation}\label{actionT}\hskip 1cm
 \delta^{(2)}S_{\rm PET}^T = 2\int{\rm d} ^3x\sqrt{g} \,\Big\{
       \big(\frac{2b_2}{\ell^2}+\beta\big)u_T^\mu\big(-\nabla^2+\frac{2}{\ell^2}\big)u_\mu^T
       -u_T^\mu\big(-\nabla^2+\frac{2}{\ell^2}\big)v_\mu^T\Big\} \,,
\end{equation}
and the scalar contribution is
\begin{align}\hskip -.8cm
 \delta^{(2)}S_{\rm PET}^{\rm scalar}=\int{\rm d} ^3x\sqrt{g}\,\Big\{ &
     \frac{\bar{\sigma}}{18}h\big(-\nabla^2+\frac{3}{\ell^2}\big)h
       -\frac19\bar{K}\big(-\nabla^2+\frac{3}{\ell^2}\big)h
     -\frac{2b_2}{9}\bar{F}\big(-\nabla^2+\frac{3}{\ell^2}\big)\bar{F} \nonumber\\
    &+ \big(\frac{2b_2}{\ell^2}+\beta\big)\big[
             -\frac23\bar{F}^2-\frac83\bar{F}\nabla^2\delta-\frac{8}{\ell^2}\delta\nabla^2\delta\big] \label{actionS}\\
    &+ \frac23\bar{F}\bar{K}+\frac43\bar{F}\nabla^2\varepsilon+\frac43\bar{K}\nabla^2\delta
             +\frac{8}{\ell^2}\delta\nabla^2\varepsilon  \Big\} \,, \nonumber
\end{align}
where $\varepsilon$ is the scalar part coming from the decomposition $v^\mu=v^\mu_T + \nabla^\mu\varepsilon$.

\subsubsection{The critical/logarithmic loci}\label{critloc}

We now evaluate the path-integrals (\ref{actionTT})--(\ref{actionS}) for the critical values of the parameters $\beta$ and
$b_2$ that lead to dual LCFTs of rank two and three.
\vspace*{,5cm}
\\
\emph{Tricritical point}\hspace*{,3cm}
Let us now consider in particular the tricritical point. At this critical locus the combinations $\bar{\sigma}$ and
$(2b_2/\ell^2+\beta)$ vanish, thus the formulas (\ref{actionTT})--(\ref{actionS}) simplify considerably. The path-integral
over the transverse-traceless tensor modes is\footnote{We integrate first over $h^{TT}$ which yields a delta function
for $k^{TT}$ and a determinant factor. The integral over $k^{TT}$ is then done trivially and we are left with an easy
integral over $f^{TT}$.
In the following we will not always try to diagonalize the action. Moreover, we will extensively make use of the delta
functions emerging from ``mixed'' path-integrals such as $h^{TT}\nabla^2k^{TT}$. For the benefit of the reader we will often denote
the order of integration.} 
\begin{equation}\label{ZTT}\hskip 1cm
 Z^{TT}_{\rm crit} = \int\mathcal{D}h_{\mu\nu}^{TT}\mathcal{D}k_{\mu\nu}^{TT}\mathcal{D}f_{\mu\nu}^{TT}\,
     e^{-\delta^{(2)}S_{\rm PET}^{TT}} = \big[\det(-\nabla^2-\frac{2}{\ell^2})_2^{TT}\big]^{-3/2} \,.
\end{equation}
To perform the integral over the transverse vector and scalar fluctuations we note that the kinetic terms have the
wrong sign. This can be remedied by a Gibbons--Hawking--Perry \cite{Gibbons:1978ac} rotation of the fluctuations
to imaginary values, which we shall also employ repeatedly in the remainder of this work. The results are
\begin{equation}\label{ZT}\hskip 1cm
 Z^T_{\rm crit}=\int\mathcal{D}v^T_\mu\mathcal{D}u^T_\mu\,e^{-\delta^{(2)}S_{\rm PET}^T}
     = \big[\det(-\nabla^2+\frac{2}{\ell^2})_1^T\big]^{-1} \,,
\end{equation}
and after integration over $h$, $\bar{K}$, $\delta$, $\varepsilon$ and finally $\bar{F}$ in that order:
\begin{equation}\hskip -.8cm
 Z^{\rm scalar}_{\rm crit} = \int\mathcal{D}h\mathcal{D}\bar{K}\mathcal{D}\varepsilon\mathcal{D}\bar{F}\mathcal{D}\delta \,
     e^{-\delta^{(2)}S_{\rm PET}^{\rm scalar}}
     = \big[\det(-\nabla^2+\frac{3}{\ell^2})_0\big]^{-3/2}\big[\det(-\nabla^2)_0\big]^{-1} \,. \label{ZS}
\end{equation}
The gravity one-loop partition function for PET gravity at the tricritical point is obtained by carefully collecting all
ghost determinants, see (\ref{Zghost}), and the determinants (\ref{ZTT}), (\ref{ZT}) and (\ref{ZS}):
\begin{equation}\begin{split}\label{Zpetdet}\hskip 1cm
 Z_{\rm crit~PET}^{\rm 1-loop}=& \,Z_{\rm gh}^3\cdot J_1^2\cdot Z^{TT}_{\rm crit}\cdot Z^T_{\rm crit}\cdot Z^{\rm scalar}_{\rm crit} \\
     =&\,\frac{\big[\det(-\nabla^2+\frac{2}{\ell^2})_1^T\big]^{1/2}}{\big[\det(-\nabla^2-\frac{2}{\ell^2})_2^{TT}\big]^{3/2}}
     = Z_{\rm Ein}\cdot \Big(\big[\det(-\nabla^2-\frac{2}{\ell^2})_2^{TT}\big]^{-1/2}\Big)^2 \,,
\end{split}\end{equation}
with $Z_{\rm Ein}$ being the one-loop contribution to the partition function of Einstein gravity, see eq.~(\ref{Zein}).
Using heat kernel techniques \cite{Giombi:2008vd,David:2009xg} it is straightforward to obtain the
result (for positive temperature [$\tau_2>0$])
\begin{equation}\label{Zpet}\hskip 1cm
 Z_{\rm crit~PET}^{\rm 1-loop}= \prod_{n=2}^\infty\frac{1}{|1-q^n|^2} \,\bigg[
                           \prod_{m=2}^\infty\prod_{\bar{m}=0}^\infty\frac{1}{1-q^m\bar{q}^{\bar{m}}}
                           \prod_{l=0}^\infty\prod_{\bar{l}=2}^\infty\frac{1}{1-q^l\bar{q}^{\bar{l}}} \bigg]^2 \,.
\end{equation}
Here we use the notation of \cite{Giombi:2008vd}, where $q=\exp(i\tau)$ with $\tau=\tau_1+i\tau_2$ and $\tau_1$ and
$\tau_2$ being related to the angular momentum $\theta$ and the inverse temperature $\beta$. \\
This can be compared to the partition function of single-particle excitations in a parity even rank-three LCFT.
\vspace*{,5cm}
\\
\emph{Single log}\hspace*{,3cm}
Now we consider the critical line in parameter space where one of the massive modes degenerates with the massless graviton,
$\beta=-3b_2/\ell^2-2\sigma\ell^2$, which still implies $\bar{\sigma}=0$. An interesting scaling limit is $b_2\to\infty$
which we will treat separately in section \ref{pmPET}.

The calculation of the transverse-traceless part is simply done and yields
\begin{equation}\label{ZlogTT}\hskip 1cm
 Z_{\rm log}^{TT}= \big[\det(-\nabla^2-\frac{2}{\ell^2})_2^{TT}\big]^{-1}
                   \big[\det(-\nabla^2-\frac{3}{\ell^2}-\frac{2\sigma\ell^2}{b_2})_2^{TT}\big]^{-1/2} \,.
\end{equation}
For the vector part we integrate over $v^T$ to obtain a delta function for $u^T$ and find the same result we obtained earlier,
see eq.~(\ref{ZT}). For the scalar part we redefine $\varepsilon=\alpha\varepsilon'$ and
$\delta'=\delta+\varepsilon'/2+\bar{F}\ell^2/6$ with $\alpha=b_2/\ell^2+2\sigma\ell^2$. These redefinitions do not produce
further ghost determinants. The point where $\alpha$ becomes zero is the tricritical point, which we already covered in the
previous subsection. After integration over $h$, $\bar{K}$, $\varepsilon'$, $\delta'$ and $\bar{F}$ we find the result (\ref{ZS}).
Collecting all contributions the result is
\begin{equation}\begin{split}\label{Zlogpetdet}\hskip 1cm
 Z_{\rm log~PET}^{\rm 1-loop}= Z_{\rm Ein}\cdot \big[\det(-\nabla^2-\frac{2}{\ell^2})_2^{TT}\big]^{-1/2}
  \big[\det(-\nabla^2-\frac{3}{\ell^2}-\frac{2\sigma\ell^2}{b_2})_2^{TT}\big]^{-1/2}\,.
\end{split}\end{equation}                                                                
Setting $b_2=-2\sigma\ell^4$ in the final expression (\ref{Zlogpetdet}), we find that it reduces to (\ref{Zpetdet}). Another
cross-check is the limit going to critical NMG, $b_2\to0$. This is not apparent from formula (\ref{Zlogpetdet}). We need to
set $b_2=0$ in the action (\ref{actionTT}) to see that there is no contribution from the integral over
$\mathcal{D}f_{\mu\nu}^{TT}$, i.e.~we do not get the last factor in (\ref{Zlogpetdet}). Therefore we obtain exactly the same
result that was obtained for critical NMG \cite{Gaberdiel:2010xv}.

For arbitrary $b_2$ ($\neq-2\sigma\ell^4$) we find that the partition function (\ref{Zlogpetdet}) consist of the contribution
of one massive mode (third term) and the contribution of critical NMG. We thus find a parity even
rank-two LCFT plus one massive mode.

The last term of eq.~(\ref{Zlogpetdet}) in terms of $q$'s is
\begin{equation}\label{ZM}\hskip 1cm
 Z_\mathcal{M}=\prod_{l=|\mathcal{M}|}^\infty\prod_{\bar{l}=|\mathcal{M}|}^\infty
          \frac{1}{(1-q^{l+1}\bar{q}^{\bar{l}-1})(1-q^{l-1}\bar{q}^{\bar{l}+1})}\,,
\end{equation}
where we defined $\mathcal{M}^2=-2\sigma\ell^2/b_2$.\footnote{Note that $\mathcal{M}^2$ defined in this way differs from
the mass squared of the propagating massive mode --- which we denote by a non-script $M^2$ --- by a factor $1/\ell^2$. Therefore,
log-modes for which $M^2=0$ have $\mathcal{M}^2=1/\ell^2$.} 
\vspace*{,5cm}
\\
\emph{Massive log}\hspace*{,3cm}
In the massive log case the parameters $\beta$ and $b_2$ are restricted by the equation
\begin{equation}\label{Mp=Mm}\hskip 1cm
 \beta^2+4b_2\sigma+\frac{6b_2\beta}{\ell^2}+\frac{10b_2^2}{\ell^4}=0 \,.
\end{equation}
Here a little more effort is needed to evaluate the 'TT' and the scalar path-integrals. The vector components do not change; we
find the result (\ref{ZT}). To evaluate the 'TT' part we first rescale $2k^{TT}_{\mu\nu}=\bar{\sigma}\bar{k}^{TT}_{\mu\nu}$ and
shift $\bar{h}^{TT}_{\mu\nu}=h^{TT}_{\mu\nu}-1/2\bar{k}^{TT}_{\mu\nu}$. Then $\bar{h}^{TT}$ decouples and we can integrate over
it. It is not possible to diagonalize in the remaining variables, but we can ``block-diagonalize'' $\bar{k}^{TT}$ and $f^{TT}$ in
the following way. Defining
\begin{equation}\hskip 1cm
 A=a\,\bar{k}^{TT}+b\,f^{TT} \qquad{\rm and}\qquad B=c\,\bar{k}^{TT}+f^{TT}
\end{equation}
we replace the remainder of the TT path-integral expression by
\begin{equation}\hskip 1cm
 A(-\nabla^2+m_A)B+B(-\nabla^2+m_B)B \,.
\end{equation}
This fixes all variables. Provided that $a$ and $b$ are not simultaneously zero the path-integral over $A$ yields a determinant
depending on $m_A$ and a delta function for $B$. We find $m_A=-2/\ell^2+M^2$, where $M^2=M_-^2=M_+^2$ is the mass of the two
propagating massive gravitons, thus
\begin{equation}\hskip 1cm
 Z_{\rm mlog}^{TT}= \big[\det(-\nabla^2-\frac{2}{\ell^2})_2^{TT}\big]^{-1/2}
                   \big[\det(-\nabla^2-\frac{2}{\ell^2}+M^2)_2^{TT}\big]^{-1} \,.
\end{equation}
To obtain the result for the scalar sector we rescale $2\bar{K}=\bar{\sigma}\bar{K}'$, $\varepsilon=\alpha\varepsilon'$ with
$\alpha$ as defined earlier, and shift $\bar{h}=h-\bar{K}'/2$, $\delta'=\delta+\ell^2/6\bar{F}$. Then we integrate over $\bar{h}$,
$\varepsilon'$ and $\delta'$ to obtain a quadratic expression in $\bar{k}$ and $\bar{F}$. This we ``block-diagonalize'' again to
find exactly (\ref{ZS}), provided $\beta\neq-2b_2/\ell^2$. Setting $\beta=-2b_2/\ell^2$, together with equation (\ref{Mp=Mm}),
implies
\begin{equation}\hskip 1cm
 (b_2,\beta)=(0,0) \qquad{\rm or }\qquad (b_2,\beta)=(-2\sigma\ell^4,4\sigma\ell^2) \,.
\end{equation}
The first solution is Einstein--Hilbert gravity while the second one is the tricritical point, so we already covered those.

The full one-loop gravity partition function for massive-log PET gravity reads
\begin{equation}\hskip 1cm
 Z_{\rm mlog~PET}^{\rm 1-loop}= Z_{\rm Einstein}\cdot (Z_\mathcal{M})^2 \,,
\end{equation}
where we defined $\mathcal{M}^2=M^2+1/\ell^2$ and $Z_\mathcal{M}$ is given in (\ref{ZM}).

\subsubsection{A special point}\label{pmPET}

The theory shows very interesting behavior if, in addition to the single log limit $\beta=-3b_2/\ell^2-2\sigma\ell^2$, we take the scaling
limit $b_2\to\infty$. In order to do so it is advised to rescale the auxiliary field $f_{\mu\nu}$. After rescaling
\begin{equation}\hskip 1cm
 \tilde{f}_{\mu\nu}=\sqrt{2b_2}\,f_{\mu\nu}\,,
\end{equation}
all fields in the linearized action (\ref{linaction}) are normalized ``canonically'', i.e.~there are no dimensionfull parameters
multiplying terms such as $(h/k/f)^{\mu\nu}\mathcal{G}_{\mu\nu}(h/k/f)$. If we now take the limit $b_2\to\infty$, keeping
$b_2/\kappa^2=1/\kappa'^2$ finite, the action (\ref{linaction}) becomes
\begin{equation}\begin{split}\label{pmaction}\hskip 1cm
 \delta^{(2)} S_{\rm pmPET} = \frac{1}{\kappa'^2}\int{\rm d} ^3x\sqrt{g} \, \Big\{
        &  k^{\mu\nu}\mathcal{G}_{\mu\nu}(h) +\tilde{f}^{\mu\nu}\mathcal{G}_{\mu\nu}(\tilde{f})
         -\frac{1}{2\ell^2}(\tilde{f}^{\mu\nu}\tilde{f}_{\mu\nu}-\tilde{f}^2)\Big\} \,.
\end{split}\end{equation}

The theory with the linearized action (\ref{pmaction}) has much more gauge symmetry then the original theory (\ref{linaction}).
First, using the self-adjointness of the tensor operator $\mathcal{G}_{\mu\nu}$ we can write the first term as
$h^{\mu\nu}\mathcal{G}_{\mu\nu}(k)$. Thus, the vector modes $v_\mu$ coming from the split of the auxiliary field $k_{\mu\nu}$,
see eq.~(\ref{tensorsplit}), are gauge modes similar to the vector part $\xi_\mu$ of $h_{\mu\nu}$. Therefore we have an
additional infinitesimal vector gauge symmetry.

Secondly, we note that the $\tilde{f}_{\mu\nu}$ terms match exactly the Lagrangian of
partially massless gravity, a certain parameter limit of NMG \cite{Bergshoeff:2009aq}. One of the massive modes becomes
partially massless in the sense of Deser and Waldron
\cite{pmrefs}
and the theory is invariant under the gauge transformation
\begin{equation}\label{infinv}\hskip 1cm
 \delta\tilde{f}_{\mu\nu}=\nabla_\mu\nabla_\nu\zeta-\frac{1}{\ell^2}g_{\mu\nu}\zeta \,,
\end{equation}
with an infinitesimal scalar gauge parameter $\zeta$. Further features of 'partially massless NMG' were discussed
in \cite{Bergshoeff:2009aq} and \cite{Grumiller:2010tj}.

However, for NMG it was shown that this gauge invariance only appears at the linearized level \cite{Blagojevic:2011qc} and
does not persist in the non-linear theory. We expect the same to be true here and it would be interesting to verify this.
Furthermore, we would like to point out the relation to the ``canonical bifurcation'' \cite{Deser:2012ci} effect taking place in the
three-dimensional pure quadratic curvature model of \cite{Deser:2009hb}. In fact, for PET gravity in the above mentioned
limit all conditions (i)--(iv) of \cite{Deser:2012ci}  for the ``canonical bifurcation'' are met.

Another remark concerns the values of the masses of the propagating modes. This is directly related to the background around
which we linearize the theory. So far we have considered our background to be AdS. In \cite{Bergshoeff:2012ev} the limit
$b_2\to\infty$ is not contained in the physical range of the parameters $\beta$ and $b_2$ because (one of) the mass squared is
negative. This is a consequence of choosing an AdS background. In analogy to 'partially massless NMG'
\cite{Bergshoeff:2009aq,Grumiller:2010tj} the mass squared of the partially massless mode in an AdS background is negative and
saturates the Breitenlohner--Freedman bound \cite{Breitenlohner:1982bm}.

Formally, it is easy to obtain (\ref{pmaction}) and (\ref{infinv}) in a de Sitter background by replacing $1/\ell^2=-\Lambda$,
with positive $\Lambda$. Then (\ref{pmaction}) does indeed propagate partially massless modes with positive mass squared.

Because the nature of the additional symmetries at the partially massless point is not clear to us, e.g.~whether they exist
also at the non-linear level, we will not discuss further the partition function of this particular theory but go on to
discuss the better understood critical points of PET gravity and their LCFT duals.

\subsection{CFT interpretation}\label{CFTinterpret}

Following the logic laid out in \cite{Gaberdiel:2010xv} and \cite{Bertin:2011jk} we present the partition functions
of the conjectured LCFT duals. Lacking a better knowledge/understanding of LCFT partition functions we give only the
partition functions corresponding to single-particle log-excitations on the CFT side. We find perfect agreement of the results
at that level.
\vspace*{,5cm}
\\
\emph{Tricritical point}\hspace*{,3cm}
In the double-log case PET gravity is conjectured to be dual to a parity even rank-three LCFT. For such a LCFT the
single-particle contribution is
\begin{equation}\label{log2CFT}\hskip 1cm
 Z_{\rm Double~log}^{\rm 1-particle} = \prod_{n=2}^\infty\frac{1}{|1-q^n|^2} \,
      \Big(1+\frac{2q^2+2\bar{q}^2}{|1-q|^2}\Big) \,.
\end{equation}
We can thus interpret the result (\ref{Zpet}) as
\begin{equation}\label{cftl2PET}\hskip 1cm
 Z_{\rm PET}^{\rm log^2}=Z_{\rm Double~log}^{\rm 1-particle}+{\rm multiparticle}\,,
\end{equation}
where the multi-particle contribution is given by
\begin{equation}\label{Zmultiparticle}\hskip 1cm
 Z_{\rm multiparticle}=\sum_{h,\bar{h}}N_{h,\bar{h}}q^h\bar{q}^{\bar{h}}\prod_{n=1}^\infty\frac{1}{|1-q^n|^2} \,.
\end{equation}
By explicit calculation of $N_{h,\bar{h}}$ for low values of $h$ and $\bar{h}$, and using the combinatorial counting
argument from \cite{Gaberdiel:2010xv,Bertin:2011jk}, one can show that all $N_{h,\bar{h}}$'s are non-negative integers.
Thus, we can interpret (\ref{Zmultiparticle}) as the contribution of physical states to the partition function.
\vspace*{,5cm}
\\
\emph{Single log}\hspace*{,3cm}
For $\beta=-3b_2/\ell^2-2\sigma\ell^2$ and $b_2\neq\pm2\sigma\ell^4$ PET gravity is conjectured to be dual to a parity even
rank-two LCFT, plus an additional massive mode. The single-particle partition function is
\begin{equation}\label{log1mCFT}\hskip 1cm
   Z_{\rm Single~log}^{\rm 1-particle} = \prod_{n=2}^\infty\frac{1}{|1-q^n|^2} \,
      \Big(1+\frac{q^2+\bar{q}^2}{|1-q|^2}+\frac{q^{|\mathcal{M}|+1}\bar{q}^{|\mathcal{M}|-1}+q^{|\mathcal{M}|-1}\bar{q}^{|\mathcal{M}|+1}}{|1-q|^2}\Big) \,.
\end{equation}
Again, we can write
\begin{equation}\label{cftlPET}\hskip 1cm
 Z_{\rm PET}^{\rm log}=Z_{\rm Single~log}^{\rm 1-particle}+Z_{\rm multiparticle} \,,
\end{equation}
with $Z_{\rm multiparticle}$ given in (\ref{Zmultiparticle}), and for given $\mathcal{M}$ show that all $N_{h,\bar{h}}$'s
are non-negative integers.
\vspace*{,5cm}
\\
\emph{Massive log}\hspace*{,3cm}
In the parameter range where $M_+=M_-=M$, PET gravity is conjectured to be dual to a parity even rank-two LCFT with non-vanishing
central charge given by
\begin{equation}\hskip 1cm
 c_{R/L}=\frac{3\ell\sigma}{G}\frac{\ell^4M^4}{1+2\ell^2M^2+2\ell^4M^4}
        =\frac{3\ell\sigma}{G}\frac{(\ell^2\mathcal{M}^2-1)^2}{1-2\ell^2\mathcal{M}^2+2\ell^4\mathcal{M}^4}\,. 
\end{equation}
The single-particle CFT partition function for such a theory takes the form
\begin{equation}\label{mlogCFT}\hskip 1cm
   Z_{\rm Massive~log}^{\rm 1-particle} = \prod_{n=2}^\infty\frac{1}{|1-q^n|^2} \,
      \Big(1+\frac{2q^{|\mathcal{M}|+1}\bar{q}^{|\mathcal{M}|-1}+2q^{|\mathcal{M}|-1}\bar{q}^{|\mathcal{M}|+1}}{|1-q|^2}\Big) \,.
\end{equation}
Again, we can write
\begin{equation}\label{cftlmPET}\hskip 1cm
 Z_{\rm PET}^{\rm mlog}=Z_{\rm Massive~log}^{\rm 1-particle}+Z_{\rm multiparticle} \,,
\end{equation}
with $Z_{\rm multiparticle}$ given in (\ref{Zmultiparticle}), and for given $\mathcal{M}$ show that all $N_{h,\bar{h}}$'s
are non-negative integers.

\section{Quasi-normal modes and partition functions of (truncated) critical gravities}\label{sec3}

In this section we address the question if there are other methods to calculate the gravity one-loop partition function
circumventing the (direct) use of heat kernel techniques. We will focus on one specific idea which relies on the work of Denef,
Hartnoll and Sachdev \cite{Denef:2009kn}, where they claim that the calculation of black hole determinants via heat kernel
techniques is equal to summing over the quasi-normal mode spectrum of the theory. Calculations in higher spin gravity lend
support to this conjecture, see e.g.~\cite{Datta:2011za}. However, the issue of boundary conditions does not play such a prominent
role there and the method of summing over quasi-normal mode spectra has not been applied to critical gravity theories yet.

One motivation to look for other-than heat kernel methods is
the apparent difference of the results obtained for the gravity one-loop partition function of log-TMG \cite{Gaberdiel:2010xv}
(TMG at the chiral point with log boundary conditions) and chiral gravity \cite{Maloney:2009ck} (TMG at the chiral point with
Brown--Henneaux boundary conditions). It was argued in \cite{Castro:2011ui} that this is an artifact of the ignorance of the AdS heat
kernel to the existence of null modes and negative energy states, which should be excluded from the
spectrum of physical modes. It was suggested to calculate a different heat kernel, one that would not
include log-modes or null states. While this seems to be a formidable, if challenging, task we will take a different approach.

It is hard to impose boundary conditions on the heat kernel, but it is very simple to do so for quasi-normal modes: choosing
the ansatz for the solution determines the boundary conditions the mode will fulfill. Here we do not refer to the boundary
conditions at the black hole horizon, which need be specified for quasi-normal modes, but to the asymptotic behavior of the
mode.

An analysis like the one we carried out in section \ref{sec2} is usually around thermal AdS, see e.g.~\cite{Giombi:2008vd}.
This is not a black hole background. However, the approach of \cite{Denef:2009kn} also holds for non black hole backgrounds,
quasi-normal modes being replaced by normal modes \cite{Denef:2009kn}. On the other hand, the results obtained in section
\ref{sec2} would not change if we would take the background to be the BTZ black hole, instead of AdS;
mainly because the BTZ is obtained by identifications of AdS \cite{Banados:1992gq}. Moreover, the thermal AdS background
used above is also the background of the Euclidean BTZ black hole, with the identification $\tau\to-1/\tau$ \cite{Giombi:2008vd}.
Therefore, we shall blithely permit ourselves to go from one background to the other, as its suits our analysis. In this
section our background will be the BTZ black hole.

Quasi-normal modes for excitations around the BTZ black hole were calculated in \cite{Birmingham:2001pj}.
In the context of critical gravity this was first done for log-TMG in \cite{Sachs:2008gt,Sachs:2008yi} and a tricritical
theory was discussed recently in \cite{Kim:2012rz}.

In the following we will calculate the one-loop partition function of two critical gravity theories in three dimensions,
NMG and PET gravity, using the conjecture of \cite{Denef:2009kn}. We confirm that at the (tri-)critical point the results
agree with earlier calculations using heat kernel techniques. Based on that
confidence we calculate different truncated one-loop partition functions by subsequently summing over different
quasi-normal mode spectra, allowing log-excitations as well as not allowing them.

We will start by recalling the partition function of Einstein gravity with a negative cosmological constant. We discuss the
contributions to the partition functions of Einstein gravity, NMG and PET gravity. Then we will define a truncation of the theory
and apply it to critical NMG and tricritical PET gravity. Finally we summarize and comment on the results we have obtained.

\subsection{Partition function of Einstein gravity}

The one-loop contribution to the partition function of Einstein gravity is (see e.g.~\cite{David:2009xg})
\begin{equation}\label{Zein}\hskip 1cm
 Z_{\rm Ein}=\sqrt{\frac{\det(-\nabla^2+2)_1^T}{\det(-\nabla^2-2)_2^{TT}}} \,.
\end{equation}
Let us briefly comment on the two terms contributing to the partition function in (\ref{Zein}).
The determinant in the numerator is due to the gauge choice and corrects the path-integral measure
(see e.g.~\cite{Gaberdiel:2010xv}), while the determinant in the denominator stems from the gauge-fixed equations of motion.
Thus, for a critical theory where we observe degeneration of the equations of motion, we expect to find a multiple of the
contribution $\det(-\nabla^2-2)_2^{TT}$ in the denominator.

According to \cite{Denef:2009kn} and \cite{Datta:2011za} the determinants are evaluated using
\begin{equation}\label{zdelta}\hskip 1cm
 \frac12\ln Z_{\Delta_s}=\ln\prod_{\kappa\geq0}^\infty|1-q^{\kappa+\Delta_s}|^{-2(\kappa+1)} \,,
\end{equation}
where $\Delta_s=1+|m_s|$ can be read off from the definitions
\begin{equation}\hskip 1cm
 \det(-\nabla^2+m^2_1-2)_1^T \quad{\rm and}\quad \det(-\nabla^2+m^2_2-3)_2^{TT} \,.
\end{equation}
Comparison with (\ref{Zein}) yields $\Delta_1=3$ and $\Delta_2=2$ to give
\begin{equation}\label{ZCFT}\hskip 1cm
 Z_{\rm Ein}=\prod_{n\geq2}^\infty\frac{1}{|1-q^n|^2}\,.
\end{equation}

This has a nice interpretation as the vacuum character of a CFT \cite{Maloney:2007ud}. In the following we will refer to
any theory of gravity whose one-loop contribution is of the form (\ref{ZCFT}) as dual to Einstein gravity or an ``ordinary''
CFT as opposed to a logarithmic CFT, but we make no further restrictions. As we will see a CFT with zero
central charge gives rise to the same character --- simply because the same modes are present in the theory --- even though
they are null modes.  

We note here that massless spin-two quasi-normal modes are often dismissed because they are pure gauge, see e.g.~\cite{Sachs:2008gt}.
But as is often the case in gravity the asymptotic behavior of the modes tells us whether it is relevant or not. The Einstein
modes are large gauge transformations and as shown by Brown and Henneaux \cite{Brown:1986nw} they are of crucial importance
as generating elements of the asymptotic symmetry group. Thus, being interested in the boundary behavior, it is logical
to include those modes and that we obtain precisely the CFT vacuum character by summing over those modes that are
responsible for non-trivial diffeomorphisms at the boundary.

However, we try to argue in this work that we can \emph{choose} to include them in the quasi-normal mode spectrum or not.
Here we do include them because they are related to non-trivial excitations in the dual conformal field theory with non-zero
central charge ($c=3\ell/2G$). In the same vein, for a critical theory we can choose not to --- or, as spelled out in
\cite{Castro:2011ui}, shall not --- include them in the spectrum if they correspond to null states that lead to zero central
charge in the dual CFT.

It was pointed out too in \cite{Castro:2011ui} that even formula (\ref{ZCFT}) is over-counting states if the central charge
is too small. Then we find null states that are multi-particle states, combinations of states which have positive norm when
considered as single-particle excitations. This lead to the restriction from (\ref{ZCFT}) to a (Virasoro) minimal model character.
Since in the following we are dealing with critical theories the central charge is always zero in our case. Therefore all
non-log-modes are null modes.

We will now go on to our main objective, the partition functions of critical gravities.

\subsection{Partition function of critical NMG and PET gravity}

Let us consider the partition function of critical NMG, which is given by \cite{Gaberdiel:2010xv}
\begin{equation}\label{ZcNMG}\hskip 1cm
  Z_{\rm cNMG}=Z_{\rm Ein}\cdot\frac{1}{\sqrt{\det(-\nabla^2-2)_2^{TT}}} \,.
\end{equation}
Not surprisingly it contains $Z_{\rm Ein}$ because all solutions to
Einstein gravity are also solutions to NMG. The second factor in (\ref{ZcNMG}) comes from the massive graviton.
Here we already took the limit to critical NMG, thus, as a consequence of the degeneration of the equations of motion,
this term coincides with the spin-two contribution of Einstein gravity.

Straightforward application of formula (\ref{zdelta}) yields
\begin{equation}\label{?ZNMG?}\hskip 1cm
 Z_{\rm cNMG} =Z_{\rm Ein}\cdot\prod_{m\geq0}^\infty|1-q^{m+2}|^{-2-2m} 
                =Z_{\rm Ein}\cdot\prod_{m\geq2}^\infty\prod_{n\geq0}^\infty\frac{1}{|1-q^{m+n}|^2} \,,
\end{equation}
which, for $q=\bar{q}$,\footnote{For the non-rotating BTZ black hole $q=\bar{q}$. This was also used to show the equality of the two
approaches in \cite{Datta:2011za}. We think that equality of the two calculations should also hold for the rotating BTZ but
it is technically more challenging to show.} perfectly agrees with the result in \cite{Gaberdiel:2010xv}. On the one hand this
is a confirmation of the conjecture of \cite{Denef:2009kn}. On the other hand, since (\ref{?ZNMG?}) is equal to the result
that was used to support the LCFT conjecture, it tells us, that, to obtain (\ref{?ZNMG?}), we
already summed over the log quasi-normal frequencies. They are equal to the frequencies of the Einstein quasi-normal modes
\cite{Sachs:2008gt,Sachs:2008yi} and correspond to the poles of the retarded correlators in the dual CFT
\cite{Birmingham:2001pj}. In the log case these are double poles \cite{Sachs:2008yi}, so we must sum over them twice. Taking a look at
(\ref{ZcNMG}) we see that we actually did count the spin-two frequencies twice simply because the (square root of the)
determinant $\det(-\nabla^2-2)_2^{TT}$ appears twice.

The partition function of tricritical PET gravity was calculated in section \ref{sec2}.
\begin{equation}\label{ZcPET}\hskip 1cm
 Z_{\rm cPET}=Z_{\rm Ein}\cdot\frac{1}{\det(-\nabla^2-2)_2^{TT}}
\end{equation}
Using formula (\ref{zdelta}) we find
\begin{equation}\label{?PET?}\hskip 1cm
 Z_{\rm cPET} =Z_{\rm Ein}\cdot\bigg(\prod_{m\geq0}^\infty|1-q^{m+2}|^{-2-2m}\bigg)^2 
                =Z_{\rm Ein}\cdot\bigg(\prod_{m\geq2}^\infty\prod_{n\geq0}^\infty\frac{1}{|1-q^{m+n}|^2}\bigg)^2 \,.
\end{equation}
For $q=\bar{q}$ this agrees with (\ref{Zpet}). Just as in (\ref{?ZNMG?}) we summed over one factor square root of
$\det(-\nabla^2-2)_2^{TT}$ for each pole of the retarded correlators. To obtain (\ref{?PET?}) we need three such factors
and indeed one finds triple poles at a tricritical point \cite{Kim:2012rz}.

Now that we identified where the separate terms in (\ref{ZcNMG}) and (\ref{ZcPET}) come from we can address the issue of
truncating the theory.

The calculations in (\ref{?ZNMG?}) and (\ref{?PET?}) suggest that the conjecture of \cite{Denef:2009kn}
holds for NMG and PET gravity at the (tri-)critical point. Based on this finding, we go on to make use of the quasi-normal mode
method and excise certain modes from the spectrum by ignoring their contribution to the partition function.
For example, in critical NMG an intriguing idea would be not to sum over the spin-two modes at all because they correspond either
to null states or negative energy states. Of course this would alter formula (\ref{zdelta}) in a drastic way:
\begin{equation}\hskip 1cm
 Z_{\Delta_2}=1 \,.
\end{equation}
In the following we will comment on the implications of such a restriction.

\subsection{Hand-picked partition functions of critical gravities}

If we do not take into account the contribution from the spin-two modes the partition function of critical NMG would become
\begin{equation}\label{toomuch}\hskip 1cm
 Z_{\text{``no modes''}}=\prod_{m\geq3}^\infty\prod_{n\geq0}^\infty |1-q^{m+n}|^2\,.
\end{equation}
Hence, we would be lead to conclude that the theory deprived of its spin-two modes is not trivial.
However, it seems likely that the correct interpretation is that here we truncated too much and the resulting
theory does not make sense.\footnote{We do not make any claims about consistency of the suggested
truncations. Rather, we calculate partition functions of possible truncations of critical theories that have been put
forward in the literature elsewhere.}

If we were to cancel only the log-modes we find, in fact, for any parity even critical gravity theory without log-modes,
\begin{equation}\hskip 1cm
 \tilde{Z}_{\rm cNMG}=\sqrt{\frac{\det(-\nabla^2+2)_1^T}{\det(-\nabla^2-2)_2^{TT}}}=Z_{\rm Ein}\,.
\end{equation}
Thus, we could conclude that critical NMG with Brown--Henneaux boundary conditions is dual to Einstein gravity. However, we know
that the only propagating modes are not only pure gauge but also null states and the dual CFT has vanishing central charge.
Therefore, according to \cite{Castro:2011ui} we would have to drop also the Einstein modes which, naively,
brings us back to (\ref{toomuch}).

Let us now consider a six-derivative theory that offers ``higher-rank'' criticality and more possibilities to truncate
the theory, PET gravity.
Truncating to the zero charge sub-sector, i.e.~imposing log boundary conditions, kills the log$^2$-modes \cite{Bergshoeff:2012ev}.
If we further dismiss the Einstein modes because they are null the partition function effectively reduces to
\begin{equation}\hskip 1cm
 \tilde{Z}_{\rm PET}=\sqrt{\frac{\det(-\nabla^2+2)_1^T}{\det(-\nabla^2-2)_2^{TT}}}=Z_{\rm Ein}\,.
\end{equation}
This suggest that tricritical gravity with the above mentioned restrictions is again dual to Einstein gravity.
The propagating degrees of freedom have zero energy but non-trivial two-point functions \cite{Bergshoeff:2012ev}.
The coexistence of the two quadratic forms, energy and correlators, yielding different results is due to a linearization
instability \cite{Apolo:2012he} and the resulting theory does not seem sensible. If we truncate further the log-modes by
imposing Brown--Henneaux boundary conditions we are again left with the infamous result (\ref{toomuch}).

\subsection{Summary}

We have shown the equivalence of the quasi-normal mode and heat kernel approaches to calculate the one-loop partition
function for critical NMG and PET gravity. We then identified the contributions to the partition function of the
different modes that on the gravity side contribute to this spectrum. This identification lead us to conclude whether
or not we should include the mode when summing over the quasi-normal mode spectrum.

We applied the truncation to two
critical theories, NMG and PET gravity, which are dual to a rank-two and a rank-three LCFT respectively. We found that
a truncation of NMG by imposing Brown--Henneaux boundary conditions yields a theory dual to an ordinary CFT.
PET gravity with log boundary conditions
and truncating the gauge modes yields another theory dual to an ordinary CFT. In the case of PET gravity, however,
the theory has a linearization instability \cite{Apolo:2012he}. So, while the higher-rank criticality of
tricritical gravity seemed to offer the possibility of a truncation to a ``sensible'' sub-sector, non-linear
calculations suggest that critical theories are either dual to LCFTs, or ordinary CFTs propagating null modes.

It might look as if we were deliberately canceling the annoying terms in the partition function. The main goal of this
work was to motivate this cancellation by identifying each determinant with its corresponding quasi-normal mode. By deciding
which mode we want to keep we are at the same time deciding which determinants actually contribute and which do not.  

We did not include parity odd theories in our discussion because one of the main ingredients, eq.~(\ref{zdelta}), only works for
parity even theories (and a non-rotating BTZ black hole background). It would be very rewarding to find an expression similar to
(\ref{zdelta}) which, unlike eq.~(\ref{?ZNMG?}), yields the critical NMG partition function from \cite{Gaberdiel:2010xv} on the
nose. Furthermore, this would allow one to consider also the case of chiral gravity, or GMG at the tricritical point.

Finally, we stress again that the relation between heat kernel and quasi-normal mode approach is a conjecture. The fact
that the results obtained using both formalisms agree made us confident to think
about the truncation procedure. A rigorous proof of the conjecture of \cite{Denef:2009kn} would not only confirm our
results but also be a strong motivation to consider the parity odd case mentioned in the previous paragraph.

\section*{Acknowledgements}

The author would like to thank Eric Bergshoeff, Sjoerd de Haan, Daniel Grumiller, Wout Merbis and Jan Rosseel for
discussions and comments on the draft. This work was supported by a grant of the Dutch Academy of Sciences (KNAW).

\appendix

\section{Multiplicity Coefficients}

To show the positivity of the multiplicity coefficients $N_{h,\bar{h}}$ we use the combinatorial counting arguments of
\cite{Gaberdiel:2010xv,Bertin:2011jk}. The arguments are given explicitly for the (single) log case in \cite{Gaberdiel:2010xv}
and for the tricritical case in \cite{Bertin:2011jk}. To proof that all multiplicity coefficients are positive the first
few coefficients have to be determined explicitly. The tables in this appendix enlist those first few coefficients for
the critical cases discussed in the main text. Table \ref{tab1} enlists the coefficients for the double log case, table
\ref{tab2} the coefficients for the single log and table \ref{tab3} gives the coefficients of the massive log case. In the
latter two tables we fixed the mass parameter to be $\mathcal{M}=2$, since $\mathcal{M}=1$ would recover the double log
case while $\mathcal{M}=0$ is a partially massless mode.

\begin{table}[!ht]
{\footnotesize
\begin{center}
 \begin{tabular}{|r|cccccccc|}
  \hline
 $\bar h=$ & 0 & 1 & 2 & 3 & 4 & 5 & 6 & 7 \\ \hline
 $h=0$: & 0 & 0 & 0 & 0 & 3 & 1 & 7 & 3 \\
 $h=1$: & 0 & 0 & 0 & 0 & 1 & 3 & 3 & 9 \\
 $h=2$: & 0 & 0 & 4 & 4 & 13 & 13 & 35 & 41 \\
 $h=3$: & 0 & 0 & 4 & 4 & 13 & 23 & 47 & 77 \\
 $h=4$: & 3 & 1 & 13 & 13 & 47 & 61 & 148 & 216 \\
 $h=5$: & 1 & 3 & 13 & 23 & 61 & 115 & 238 & 422 \\
 $h=6$: & 7 & 3 & 35 & 47 & 148 & 238 & 550 & 908 \\
 $h=7$: & 3 & 9 & 41 & 77 & 216 & 422 & 908 & 1690 \\
\hline
 \end{tabular}
 \caption{Double log multiplicity coefficients $N_{h,\bar h}$ in eq.~(\ref{cftl2PET}) for $h, \bar h < 8$.}\label{tab1}
\end{center}
}
\end{table}

\begin{table}[!ht]
{\footnotesize
\begin{center}
 \begin{tabular}{|r|cccccccc|}
  \hline
 $\bar h=$ & 0 & 1 & 2 & 3 & 4 & 5 & 6 & 7 \\ \hline
 $h=0$: & 0 & 0 & 0 & 0 & 1 & 0 & 2 & 0 \\
 $h=1$: & 0 & 0 & 0 & 0 & 0 & 2 & 1 & 4 \\
 $h=2$: & 0 & 0 & 1 & 1 & 3 & 3 & 8 & 7 \\
 $h=3$: & 0 & 0 & 1 & 3 & 4 & 9 & 12 & 22 \\
 $h=4$: & 1 & 0 & 3 & 4 & 11 & 14 & 31 & 41 \\
 $h=5$: & 0 & 2 & 3 & 9 & 14 & 31 & 49 & 91 \\
 $h=6$: & 2 & 1 & 8 & 12 & 31 & 49 & 104 & 159 \\
 $h=7$: & 0 & 4 & 7 & 22 & 41 & 91 & 159 & 302 \\
\hline
 \end{tabular}
 \caption{Single log multiplicity coefficients $N_{h,\bar h}$ in eq.~(\ref{cftlPET}) for $h, \bar h < 8$
          and $\mathcal{M}=2$.}\label{tab2}
\end{center}
}
\end{table}

\begin{table}[!ht]
{\footnotesize
\begin{center}
 \begin{tabular}{|r|cccccccc|}
  \hline
 $\bar h=$ & 0 & 1 & 2 & 3 & 4 & 5 & 6 & 7 \\ \hline
 $h=0$: & 0 & 0 & 0 & 0 & 0 & 0 & 0 & 0 \\
 $h=1$: & 0 & 0 & 0 & 0 & 0 & 0 & 0 & 0 \\
 $h=2$: & 0 & 0 & 0 & 0 & 0 & 0 & 3 & 1 \\
 $h=3$: & 0 & 0 & 0 & 0 & 0 & 0 & 1 & 3 \\
 $h=4$: & 0 & 0 & 0 & 0 & 4 & 4 & 7 & 5 \\
 $h=5$: & 0 & 0 & 0 & 0 & 4 & 4 & 5 & 13 \\
 $h=6$: & 0 & 0 & 3 & 1 & 7 & 5 & 10 & 14 \\
 $h=7$: & 0 & 0 & 1 & 3 & 5 & 13 & 14 & 38 \\
\hline
 \end{tabular}
 \caption{Massive log multiplicity coefficients $N_{h,\bar h}$ in eq.~(\ref{cftlmPET}) for $h, \bar h < 8$
          and $\mathcal{M}=2$.}\label{tab3}
\end{center}
}
\end{table}


\begin{thebibliography}{10}

\bibitem{Deser:1981wh}
S.~Deser, R.~Jackiw, and S.~Templeton, ``{Topologically Massive Gauge Theories},''
 \href{http://dx.doi.org/10.1016/0003-4916(82)90164-6}{{\em Annals Phys.} {\bfseries 140} (1982) 372--411},
 \href{http://dx.doi.org/10.1016/0003-4916(88)90053-X}{Erratum-ibid.~{\bfseries 185} (1988) 406},
 \href{http://dx.doi.org/10.1016/0003-4916(88)90053-X}{{\em Annals Phys.} {\bfseries 185} (1988) 406},
 \href{http://dx.doi.org/10.1006/aphy.2000.6013}{{\em Annals Phys.} {\bfseries 281} (2000) 409--449}
 \href{http://inspirehep.net/record/169285}{\textsc{[inSPIRE]}}.

\bibitem{Bergshoeff:2009hq}
E.~A. Bergshoeff, O.~Hohm, and P.~K. Townsend, ``{Massive Gravity in Three
  Dimensions},'' \href{http://dx.doi.org/10.1103/PhysRevLett.102.201301}{{\em
  Phys.Rev.Lett.} {\bfseries 102} (2009) 201301}
\href{http://inspirehep.net/record/810887}{\textsc{[inSPIRE]}}
\href{http://arxiv.org/abs/0901.1766}{{\ttfamily [arXiv:0901.1766]}}.

\bibitem{Li:2008dq}
W.~Li, W.~Song, and A.~Strominger, ``{Chiral Gravity in Three Dimensions},''
  \href{http://dx.doi.org/10.1088/1126-6708/2008/04/082}{{\em JHEP} {\bfseries
  0804} (2008) 082}
\href{http://inspirehep.net/record/778412}{\textsc{[inSPIRE]}}
\href{http://arxiv.org/abs/0801.4566}{{\ttfamily [arXiv:0801.4566]}}.

\bibitem{Grumiller:2008qz}
D.~Grumiller and N.~Johansson, ``{Instability in cosmological topologically
  massive gravity at the chiral point},''
  \href{http://dx.doi.org/10.1088/1126-6708/2008/07/134}{{\em JHEP} {\bfseries
  0807} (2008) 134}
\href{http://inspirehep.net/record/786091}{\textsc{[inSPIRE]}}
\href{http://arxiv.org/abs/0805.2610}{{\ttfamily [arXiv:0805.2610]}}.

\bibitem{Bergshoeff:2009aq}
E.~A. Bergshoeff, O.~Hohm, and P.~K. Townsend, ``{More on Massive 3D
  Gravity},'' \href{http://dx.doi.org/10.1103/PhysRevD.79.124042}{{\em
  Phys.Rev.} {\bfseries D79} (2009) 124042}
\href{http://inspirehep.net/record/819750}{\textsc{[inSPIRE]}}
\href{http://arxiv.org/abs/0905.1259}{{\ttfamily [arXiv:0905.1259]}}.

\bibitem{Bergshoeff:2012ev}
E.~A. Bergshoeff, S.~de~Haan, W.~Merbis, J.~Rosseel, and T.~Zojer,
  ``{Three-Dimensional Tricritical Gravity},''
  \href{http://dx.doi.org/10.1103/PhysRevD.86.064037}{{\em Phys.Rev.}
  {\bfseries D86} (2012) 064037}
\href{http://inspirehep.net/record/1118305}{\textsc{[inSPIRE]}}
\href{http://arxiv.org/abs/1206.3089}{{\ttfamily [arXiv:1206.3089]}}.

\bibitem{Grumiller:2010tj}
D.~Grumiller, N.~Johansson, and T.~Zojer, ``{Short-cut to new anomalies in
  gravity duals to logarithmic conformal field theories},''
  \href{http://dx.doi.org/10.1007/JHEP01(2011)090}{{\em JHEP} {\bfseries 1101}
  (2011) 090}
\href{http://inspirehep.net/record/873622}{\textsc{[inSPIRE]}}
\href{http://arxiv.org/abs/1010.4449}{{\ttfamily [arXiv:1010.4449]}}.

\bibitem{Gaberdiel:2010xv}
M.~R. Gaberdiel, D.~Grumiller, and D.~Vassilevich, ``{Graviton 1-loop partition
  function for 3-dimensional massive gravity},''
  \href{http://dx.doi.org/10.1007/JHEP11(2010)094}{{\em JHEP} {\bfseries 1011}
  (2010) 094}
\href{http://inspirehep.net/record/863609}{\textsc{[inSPIRE]}}
\href{http://arxiv.org/abs/1007.5189}{{\ttfamily [arXiv:1007.5189]}}.

\bibitem{Bertin:2011jk}
M.~Bertin, D.~Grumiller, D.~Vassilevich, and T.~Zojer, ``{Generalised massive
  gravity one-loop partition function and AdS/(L)CFT},''
  \href{http://dx.doi.org/10.1007/JHEP06(2011)111}{{\em JHEP} {\bfseries 1106}
  (2011) 111}
\href{http://inspirehep.net/record/894389}{\textsc{[inSPIRE]}}
\href{http://arxiv.org/abs/1103.5468}{{\ttfamily [arXiv:1103.5468]}}.

\bibitem{Brown:1986nw}
J.~D. Brown and M.~Henneaux, ``{Central Charges in the Canonical Realization of
  Asymptotic Symmetries: An Example from Three-Dimensional Gravity},''
\href{http://dx.doi.org/10.1007/BF01211590}{{\em Commun.Math.Phys.} {\bfseries
  104} (1986) 207--226}
\href{http://inspirehep.net/record/231928}{\textsc{[inSPIRE]}}.

\bibitem{Grumiller:2008es}
D.~Grumiller and N.~Johansson, ``{Consistent boundary conditions for
  cosmological topologically massive gravity at the chiral point},''
  \href{http://dx.doi.org/10.1142/S0218271808014096}{{\em Int.J.Mod.Phys.}
  {\bfseries D17} (2008) 2367--2372}
\href{http://inspirehep.net/record/793388}{\textsc{[inSPIRE]}}
\href{http://arxiv.org/abs/0808.2575}{{\ttfamily [arXiv:0808.2575]}}.

\newpage

\bibitem{Bergshoeff:2012sc}
E.~A. Bergshoeff, S.~de~Haan, W.~Merbis, M.~Porrati, and J.~Rosseel, ``{Unitary
  Truncations and Critical Gravity: a Toy Model},''
  \href{http://dx.doi.org/10.1007/JHEP04(2012)134}{{\em JHEP} {\bfseries 1204}
  (2012) 134}
\href{http://inspirehep.net/record/1083409}{\textsc{[inSPIRE]}}
\href{http://arxiv.org/abs/1201.0449}{{\ttfamily [arXiv:1201.0449]}}.

\bibitem{Apolo:2012he}
L.~Apolo and M.~Porrati, ``{Nonlinear Dynamics of Parity-Even Tricritical
  Gravity in Three and Four Dimensions},''
  \href{http://dx.doi.org/10.1007/JHEP08(2012)051}{{\em JHEP} {\bfseries 1208}
  (2012) 051}
\href{http://inspirehep.net/record/1119432}{\textsc{[inSPIRE]}}
\href{http://arxiv.org/abs/1206.5231}{{\ttfamily [arXiv:1206.5231]}}.

\bibitem{Denef:2009kn}
F.~Denef, S.~A. Hartnoll, and S.~Sachdev, ``{Black hole determinants and
  quasinormal modes},''
  \href{http://dx.doi.org/10.1088/0264-9381/27/12/125001}{{\em
  Class.Quant.Grav.} {\bfseries 27} (2010) 125001}
\href{http://inspirehep.net/record/829077}{\textsc{[inSPIRE]}}
\href{http://arxiv.org/abs/0908.2657}{{\ttfamily [arXiv:0908.2657]}}.

\bibitem{Banados:1992wn}
M.~Banados, C.~Teitelboim, and J.~Zanelli, ``{The Black hole in
  three-dimensional space-time},''
  \href{http://dx.doi.org/10.1103/PhysRevLett.69.1849}{{\em Phys.Rev.Lett.}
  {\bfseries 69} (1992) 1849--1851}
\href{http://inspirehep.net/record/32290}{\textsc{[inSPIRE]}}
\href{http://arxiv.org/abs/hep-th/9204099}{{\ttfamily [arXiv:hep-th/9204099]}}.

\bibitem{Gibbons:1978ac}
G.~Gibbons, S.~Hawking, and M.~Perry, ``{Path Integrals and the Indefiniteness
  of the Gravitational Action},''
\href{http://dx.doi.org/10.1016/0550-3213(78)90161-X}{{\em Nucl.Phys.}
  {\bfseries B138} (1978) 141}
\href{http://inspirehep.net/record/6506}{\textsc{[inSPIRE]}}.

\bibitem{Giombi:2008vd}
S.~Giombi, A.~Maloney, and X.~Yin, ``{One-loop Partition Functions of 3D
  Gravity},'' \href{http://dx.doi.org/10.1088/1126-6708/2008/08/007}{{\em JHEP}
  {\bfseries 0808} (2008) 007}
\href{http://inspirehep.net/record/783201}{\textsc{[inSPIRE]}}
\href{http://arxiv.org/abs/0804.1773}{{\ttfamily [arXiv:0804.1773]}}.

\bibitem{David:2009xg}
J.~R. David, M.~R. Gaberdiel, and R.~Gopakumar, ``{The Heat Kernel on AdS(3)
  and its Applications},''
  \href{http://dx.doi.org/10.1007/JHEP04(2010)125}{{\em JHEP} {\bfseries 1004}
  (2010) 125}
\href{http://inspirehep.net/record/838200}{\textsc{[inSPIRE]}}
\href{http://arxiv.org/abs/0911.5085}{{\ttfamily [arXiv:0911.5085]}}.

\bibitem{pmrefs}
S.~Deser and R.~I. Nepomechie, ``{Gauge invariance versus masslessness in de
  Sitter space},''
\href{http://dx.doi.org/10.1016/0003-4916(84)90156-8}{{\em Annals Phys.}
  {\bfseries 154} (1984) 396}
\href{http://inspirehep.net/record/13830}{\textsc{[inSPIRE]}}.

S.~Deser and A.~Waldron, ``{Gauge invariances and phases of massive higher
  spins in (A)dS},''
  \href{http://dx.doi.org/10.1103/PhysRevLett.87.031601}{{\em Phys.Rev.Lett.}
  {\bfseries 87} (2001) 031601}
\href{http://inspirehep.net/record/553479}{\textsc{[inSPIRE]}}
\href{http://arxiv.org/abs/hep-th/0102166}{{\ttfamily [arXiv:hep-th/0102166]}}.

S.~Deser and A.~Waldron, ``{Partial masslessness of higher spins in (A)dS},''
  \href{http://dx.doi.org/10.1016/S0550-3213(01)00212-7}{{\em Nucl.Phys.}
  {\bfseries B607} (2001) 577--604}
\href{http://inspirehep.net/record/554444}{\textsc{[inSPIRE]}}
\href{http://arxiv.org/abs/hep-th/0103198}{{\ttfamily [arXiv:hep-th/0103198]}}.

\bibitem{Blagojevic:2011qc}
M.~Blagojevic and B.~Cvetkovic, ``{Extra gauge symmetries in BHT gravity},''
  \href{http://dx.doi.org/10.1007/JHEP03(2011)139}{{\em JHEP} {\bfseries 1103}
  (2011) 139}
\href{http://inspirehep.net/record/892526}{\textsc{[inSPIRE]}}
\href{http://arxiv.org/abs/1103.2388}{{\ttfamily [arXiv:1103.2388]}}.

\bibitem{Deser:2012ci}
S.~Deser, S.~Ertl, and D.~Grumiller, ``{Canonical bifurcation in higher
  derivative, higher spin, theories},''
\href{http://inspirehep.net/record/1125560}{\textsc{[inSPIRE]}}
\href{http://arxiv.org/abs/1208.0339}{{\ttfamily [arXiv:1208.0339]}}.

\bibitem{Deser:2009hb}
S.~Deser, ``{Ghost-free, finite, fourth order D=3 (alas) gravity},''
  \href{http://dx.doi.org/10.1103/PhysRevLett.103.101302}{{\em Phys.Rev.Lett.}
  {\bfseries 103} (2009) 101302}
\href{http://inspirehep.net/record/818959}{\textsc{[inSPIRE]}}
\href{http://arxiv.org/abs/0904.4473}{{\ttfamily [arXiv:0904.4473]}}.

\bibitem{Breitenlohner:1982bm}
P.~Breitenlohner and D.~Z. Freedman, ``{Positive Energy in anti-De Sitter
  Backgrounds and Gauged Extended Supergravity},''
\href{http://dx.doi.org/10.1016/0370-2693(82)90643-8}{{\em Phys.Lett.}
  {\bfseries B115} (1982) 197}
\href{http://inspirehep.net/record/12129}{\textsc{[inSPIRE]}}.

\bibitem{Datta:2011za}
S.~Datta and J.~R. David, ``{Higher Spin Quasinormal Modes and One-Loop
  Determinants in the BTZ black Hole},''
  \href{http://dx.doi.org/10.1007/JHEP03(2012)079}{{\em JHEP} {\bfseries 1203}
  (2012) 079}
\href{http://inspirehep.net/record/1082354}{\textsc{[inSPIRE]}}
\href{http://arxiv.org/abs/1112.4619}{{\ttfamily [arXiv:1112.4619]}}.

\bibitem{Maloney:2009ck}
A.~Maloney, W.~Song, and A.~Strominger, ``{Chiral Gravity, Log Gravity and
  Extremal CFT},'' \href{http://dx.doi.org/10.1103/PhysRevD.81.064007}{{\em
  Phys.Rev.} {\bfseries D81} (2010) 064007}
\href{http://inspirehep.net/record/816360}{\textsc{[inSPIRE]}}
\href{http://arxiv.org/abs/0903.4573}{{\ttfamily [arXiv:0903.4573]}}.

\bibitem{Castro:2011ui}
A.~Castro, T.~Hartman, and A.~Maloney, ``{The Gravitational Exclusion Principle
  and Null States in Anti-de Sitter Space},''
  \href{http://dx.doi.org/10.1088/0264-9381/28/19/195012}{{\em
  Class.Quant.Grav.} {\bfseries 28} (2011) 195012}
\href{http://inspirehep.net/record/920289}{\textsc{[inSPIRE]}}
\href{http://arxiv.org/abs/1107.5098}{{\ttfamily [arXiv:1107.5098]}}.

\bibitem{Banados:1992gq}
M.~Banados, M.~Henneaux, C.~Teitelboim, and J.~Zanelli, ``{Geometry of the
  (2+1) black hole},'' \href{http://dx.doi.org/10.1103/PhysRevD.48.1506}{{\em
  Phys.Rev.} {\bfseries D48} (1993) 1506--1525}
\href{http://inspirehep.net/record/343161}{\textsc{[inSPIRE]}}
\href{http://arxiv.org/abs/gr-qc/9302012}{{\ttfamily [arXiv:gr-qc/9302012]}}.

\bibitem{Birmingham:2001pj}
D.~Birmingham, I.~Sachs, and S.~N. Solodukhin, ``{Conformal field theory
  interpretation of black hole quasinormal modes},''
  \href{http://dx.doi.org/10.1103/PhysRevLett.88.151301}{{\em Phys.Rev.Lett.}
  {\bfseries 88} (2002) 151301}
\href{http://inspirehep.net/record/567938}{\textsc{[inSPIRE]}}
\href{http://arxiv.org/abs/hep-th/0112055}{{\ttfamily [arXiv:hep-th/0112055]}}.

\bibitem{Sachs:2008gt}
I.~Sachs and S.~N. Solodukhin, ``{Quasi-Normal Modes in Topologically Massive
  Gravity},'' \href{http://dx.doi.org/10.1088/1126-6708/2008/08/003}{{\em JHEP}
  {\bfseries 0808} (2008) 003}
\href{http://inspirehep.net/record/787864}{\textsc{[inSPIRE]}}
\href{http://arxiv.org/abs/0806.1788}{{\ttfamily [arXiv:0806.1788]}}.

\bibitem{Sachs:2008yi}
I.~Sachs, ``{Quasi-Normal Modes for Logarithmic Conformal Field Theory},''
  \href{http://dx.doi.org/10.1088/1126-6708/2008/09/073}{{\em JHEP} {\bfseries
  0809} (2008) 073}
\href{http://inspirehep.net/record/790551}{\textsc{[inSPIRE]}}
\href{http://arxiv.org/abs/0807.1844}{{\ttfamily [arXiv:0807.1844]}}.

\bibitem{Kim:2012rz}
Y.-W. Kim, Y.~S. Myung, and Y.-J. Park, ``{Quasinormal modes around the BTZ
  black hole at the tricritical generalized massive gravity},''
  \href{http://dx.doi.org/10.1103/PhysRevD.86.064017}{{\em Phys.Rev.}
  {\bfseries D86} (2012) 064017}
\href{http://inspirehep.net/record/1122536}{\textsc{[inSPIRE]}}
\href{http://arxiv.org/abs/1207.3149}{{\ttfamily [arXiv:1207.3149]}}.

\bibitem{Maloney:2007ud}
A.~Maloney and E.~Witten, ``{Quantum Gravity Partition Functions in Three
  Dimensions},'' \href{http://dx.doi.org/10.1007/JHEP02(2010)029}{{\em JHEP}
  {\bfseries 1002} (2010) 029}
\href{http://inspirehep.net/record/769256}{\textsc{[inSPIRE]}}
\href{http://arxiv.org/abs/0712.0155}{{\ttfamily [arXiv:0712.0155]}}.

\end{thebibliography}

\newpage

\section*{References}

\providecommand{\href}[2]{#2}\begingroup\raggedright\endgroup

\end{document}